\documentclass[pra,aps,twocolumn,amsmath,amssymb,superscriptaddress]{revtex4-1}
\usepackage{epsfig,amsmath}
\usepackage{subfigure}
\usepackage{graphicx}
\usepackage{dcolumn}
\usepackage{stmaryrd}
\usepackage{mathrsfs}
\usepackage{pifont}
\usepackage{amsthm}
\usepackage{amssymb}
\usepackage{bm}
\usepackage{latexsym}
\usepackage[colorlinks=true,linkcolor=blue,citecolor=blue]{hyperref}
\usepackage{color}
\usepackage{epstopdf}
\usepackage[ruled]{algorithm2e}

\begin{document}

\title{An almost deterministic cooling by measurements}

\author{Jia-shun Yan}
\affiliation{School of Physics, Zhejiang University, Hangzhou 310027, Zhejiang, China}

\author{Jun Jing}
\email{Email address: jingjun@zju.edu.cn}
\affiliation{School of Physics, Zhejiang University, Hangzhou 310027, Zhejiang, China}

\date{\today}

\begin{abstract}
Nondeterministic measurement-based techniques are efficient in reshaping the population distribution of a quantum system but suffer from a limited success probability of holding the system in the target state. To reduce the experimental cost, we exploit the state-engineering mechanisms of both conditional and unconditional measurements and propose a two-step protocol assisted by a qubit to cool a resonator down to the ground state with a near-unit probability. In the first step, the unconditional measurements on the ancillary qubit are applied to reshape the target resonator from a thermal state to a reserved Fock state. The measurement sequence is optimized by reinforcement learning for a maximum fidelity. In the second step, the population on the reserved state can be faithfully transferred in a stepwise way to the resonator's ground state with a near-unit fidelity by the conditional measurements on the qubit. Intrinsic nondeterminacy of the projection-based conditional measurement is effectively inhibited by properly spacing the measurement sequence, which makes the Kraus operator act as a lowering operator for neighboring Fock states. Through dozens of measurements, the initial thermal average occupation of the resonator can be reduced by five orders in magnitude with a success probability over $95\%$.
\end{abstract}

\maketitle

\section{Introduction}

Microscopic and mesoscopic resonators exhibit nonclassical behaviors when they are cooled nearly down to the ground states. As a crucial prerequisite for initialization of a quantum system~\cite{SpinQuantumRegister,TenQubitRegister}, adiabatic quantum computing~\cite{AdiabaticQC,QuantumAnnealing}, and ultrahigh-precision measurements~\cite{QNDMeasurement,ContinuousMeasurement}, the ground-state cooling has attracted numerous interests in recent decades~\cite{LaserCooling,LaserCooling2, LaserCooling3,CoolingMagnon,SidebandCooling,SidebandCooling2016}. Particularly with the laser technique, the interaction established between the resonator (as an external degree of freedom of an atomic or molecular system) and the spin (as an inner degree of freedom) provides a decay channel or an asymmetric transition for the energy leakage of the resonator. Rich physics can be discovered on mechanical ground-state preparation in the presence~\cite{FeedbackCooling} or in the absence of feedback control~\cite{CoolingWithoutFeedback}. A relevant yet profoundly distinct method takes advantage of the quantum measurement in the same setting of resonator-spin interaction. It shows a dramatic efficiency and works as a powerful tool for quantum computation~\cite{OneWayComputer,MeasurementVQE,MeasurementComputation,51QubitsMeasurementVQE}, quantum state preparation~\cite{OptimizedPreparation,MeasrementEvolution,CoolingByPulses,CoolingByPulsesExp,MeasurementCharging} and entanglement transition~\cite{ZenoEntanglementTransition,EntanglementDynamics,PhaseTransition}.

Deterministic and nondeterministic protocols on measurement-based cooling roughly constitute two main branches, depending on whether the cooling procedure is unconditionally continued or not. Feedback loops are required in many protocols of deterministic cooling. According to the optical readout of the position information of the mechanical resonator, the controls over interpulse spacing~\cite{CoolingByPulses,StroboscopicControl}, pulse duration~\cite{CoolingByPulses}, and exerted force~\cite{MotionMeasurementControl} are carried out to realize refrigeration. In nondeterministic cooling, the cooling loops are probabilistically catenated~\cite{Li2011,Cooling2009,Purification,Nonlinear}. Only upon an outcome of projective measurement implying that the measured system is in the target state, the cooling process is continued. Otherwise, the system sample is abandoned and the whole process is restarted. Projection induced sequential postselections gradually force the mechanical oscillator into its ground state via dynamically filtering out its vibrational modes~\cite{MeasCoolingExp,DemonlikeCooling}. Through actively selecting the desired results, the nondeterministic cooling is overwhelmingly efficient in the average-population-reduction rate, which greatly reduces the number of cooling loops. However, its success probability, the product of measurement probabilities of postselections, is limited as a cost of the high efficiency. For a cooling protocol entirely consisting of conditional measurements, the success probability is usually in the order of $10\%$ or less. For example, a measurement-based cooling of a mechanical resonator was proposed in Ref~\cite{Li2011}, where the average phonon number is reduced by four orders in magnitude with a success probability about $20\%$. In Ref~\cite{Nonlinear}, the average phonon number of a nonlinear mechanical resonator is reduced by about eight orders in magnitude with a success probability about $10\%$. In Ref.~\cite{MeasCoolingExp}, a trapped ion oscillator is cooled down to nearly a ground state by measurement-based cooling with a success probability about $12\%$. Undoubtedly, a low success probability would raise many problems in practice and add extra complications in experiments. Therefore, the success probability should be addressed as a principle element for evaluating nondeterministic cooling protocols.

To improve the success probability, one can reduce the number of projections~\cite{OneShotMeasurement,OptimizedCooling,Li2011,CoolingWithoutFeedback} or use by-product operators and adaptive measurements~\cite{MeasurementVQE,MQCClusterState}. An alternative yet surprisingly unexplored idea might be purifying the target system before performing the projective measurements. We employ the unconditional or nonselective measurement~\cite{VonNeumann1955Mathematical,ControlvonNeuman} by virtue of its high capacity in population concentration from low-energy states to certain reserved high-energy states. The unconditional measurement is characterized by a pure dephasing operation with partial collapse of the wave-function~\cite{UnconditionalExp} and it has no preference for projecting the measured system onto a subspace~\cite{FockState}. It is therefore a promising tool to realize state engineering with intact populations. After the target system is effectively purified, its dynamics becomes more predictable, making it more likely to be manipulated by projective measurements with a near-unit success probability.

In this paper, we propose a two-step cooling framework based on measurements. In the first step, unconditional measurements are constantly performed on an ancillary qubit prepared as the excited state to gradually transform the coupled resonator from a thermal state to a reserved Fock state. Measurements with a shorter interval tend to collect more populations from other states to the reserved state, while those with a larger interval are inclined to sharpen the state distribution around the reserved state. The optimized time-spacing sequence, that is desired to reshape a thermal state to an almost pure state, can be generated by the reinforcement learning. It is a powerful tool to learn complex behaviours directly from reward signals in both classical~\cite{GameGoNN,GameGoWithoutHuman,GameChess,HumanLevelControl} and quantum systems or environments~\cite{MLAndPhysics,ErrorCorrectionFeedback,DigitalQuantumSimulation,StatePreparationRL}. In our second step, projective measurements are utilized to steer the reserved Fock state back to the mechanical ground state by stepwise positive operator-valued measures (POVM). The POVM is generated by measuring the excited state of the ancillary qubit that is prepared in ground state, yielding a near-prefect population transfer between neighboring Fock states of the resonator. With updatable measurement intervals in an analytical formula, a mechanical resonator is cooled down to the ground state with a success probability over $95\%$.

The rest part of this paper is structured as follows. In Sec.~\ref{ModelAndMeaSec}, we introduce a framework for pure state preparation based on unconditional measurements and population transfer induced by conditional measurements. Analytically we provide the conditions for reserving a proper state and the updating optimal interval for the conditional measurements. In Sec.~\ref{UnitProbCoolingSec}, we present the two-step cooling protocol and demonstrate the cooling dynamics of the mechanical resonator under measurements. In Sec.~\ref{discussion}, we study the robustness of the ground-state fidelity and the success probability for various reserved states against the thermal decoherence. We summarize our work in Sec.~\ref{ConclusionSec}.

\section{Model and measurements}\label{ModelAndMeaSec}

Both unconditional and conditional measurements in our two-step cooling protocol are based on the Jaynes-Cummings (JC) model. For simplicity and with no loss of generality, the ground-state energy of the ancillary qubit is set as $\omega_g=0$. The full Hamiltonian in the rotating frame with respect to $H_0=\omega_b(|e\rangle\langle e|+b^\dagger b)$ then reads ($\hbar\equiv1$)
\begin{equation}\label{Ham}
H=\Delta|e\rangle\langle e|+g(b^\dagger\sigma_-+b\sigma_+).
\end{equation}
Here $\Delta\equiv\omega_e-\omega_b$ represents the detuning between the energy splitting of the ancillary qubit $\omega_e$ and the frequency of the target resonator $\omega_b$. $g$ is the coupling strength of the JC interaction. $b$ $(b^\dagger)$ is the annihilation (creation) operator of the resonator and $\sigma_-=|g\rangle\langle e|$ and $\sigma_+=|e\rangle\langle g|$ are the transition operators of the qubit.

The resonator is assumed to be initially in a thermal bath and then has an initial state~\cite{IntroductoryQO} ($k_B\equiv1$)
\begin{equation}
\rho_b^{\rm th}=\frac{1}{1+\bar{n}_{\rm th}}\sum_{n=0}^{\infty}\left(\frac{\bar{n}_{\rm th}}{1+\bar{n}_{\rm th}}\right)^n|n\rangle\langle n|,
\end{equation}
where $\bar{n}_{\rm th}={\rm Tr}[\hat{n}\rho_b^{\rm th}]=1/(e^{\omega_b/T}-1)$ is the thermal average population and $T$ represents the temperature of the thermal bath attached to the resonator. Instantaneous von Neuman quantum measurements can be divided into two types~\cite{ControlvonNeuman}: conditional and unconditional measurements, depending on whether the measurement outcome is recorded or not. A general measurement operator could be defined by $Q=\sum_iq_iM_i$ and $QM_i=q_iM_i$ with the projector $M_i$ indicating a particular subspace. For conditional measurement~\cite{MeasurementEigensolver,MeasurementDriving,SpinBathPurification}, the state after a measurement becomes $M_i\rho M_i/{\rm Tr}[M_i\rho]$, where $\rho$ is the density matrix of the composite system of resonator and qubit. For unconditional measurement, it involves the whole space of the measured system and the state after measurement will be $\sum_iM_i\rho M_i$. In our almost-deterministic cooling protocol, both measurements are employed but for different purposes. Unconditional measurements can generate a Fock state of high energy from a thermal state, while conditional measurements are used to transfer the high-level population to its lower-energy neighbors.

\subsection{Fock-state preparation based on unconditional measurement}\label{Fockstate}

Conserving the excitation number of the whole system, the JC Hamiltonian~(\ref{Ham}) is block diagonal in the Hilbert space and could be written as
\begin{equation}
H^{(n)}=\begin{pmatrix}0&g\sqrt{n}\\g\sqrt{n}&\Delta\end{pmatrix}
\end{equation}
in the $n$-excitation subspace spanned by $\{|g,n\rangle, |e,n-1\rangle\}$. Then the time-evolution operator reads,
\begin{equation}
U=\bigoplus_n e^{-i\frac{\Delta \tau}{2}}
\left(\begin{array}{cc}\alpha_n & \beta_n\\ \beta_n & \alpha_n^*\end{array}\right),
\end{equation}
where
\begin{eqnarray}\label{AlphaBeta}
&\alpha_n=\cos(\Omega_n\tau)+i\Delta\sin(\Omega_n\tau)/2\Omega_n, \\
&\beta_n =-ig\sqrt{n}\sin(\Omega_n\tau)/\Omega_n
\end{eqnarray}
are cooling coefficients and $\Omega_n=\sqrt{g^2n+\Delta^2/4}$ is the Rabi frequency. Starting from an arbitrary mixed state $\rho_b=\sum_np_n|n\rangle\langle n|$ of the resonator and the excited state of the qubit, an unconditional measurement preformed on the qubit after a joint evolution of a period of $\tau$ can yield a superoperation $\mathcal{U}(\tau)[\rho_b]\equiv{\rm Tr}_a[U(|e\rangle\langle e|\otimes\rho_b)U^\dagger]$ and the resulting state of resonator reads
\begin{equation}\label{mathcalU}
\rho_b'(\tau)=\mathcal{U}(\tau)[\rho_b]=\sum_{n\geq0}\left(p_n|\alpha_{n+1}|^2+p_{n-1}|\beta_n|^2\right)|n\rangle\langle n|,
\end{equation}
where $p_{-1}$ is set as zero for a compact formula. Here we omit the time of performing measurement~\cite{SpinQubitMeasurement} since it is much shorter than a typical evolution period $\tau$. In regard to the population distribution of the resonator, a transfer therefore occurs between each Fock state and its lower-energy neighbor
\begin{equation}\label{pn}
p_n\rightarrow p_n|\alpha_{n+1}|^2+p_{n-1}|\beta_n|^2,
\end{equation}
where the changing amount of the population is determined by the cooling coefficients $|\alpha_{n+1}|^2$ and $|\beta_n|^2$. In particular, $|\alpha_{n+1}|^2$ implies the to-be-reserved proportion of the original population on the $n$th Fock state and $|\beta_n|^2$ acts as the weighting factor for the population on the $(n-1)$th state transferred to its upper state. The populations over certain states $|n\rangle$ would keep growing under repeated unconditional measurements on the qubit with a fixed interval that satisfies $|\alpha_{n+1}|^2=1$, since then the populations on their lower-energy neighbors are transferred to them by $p_{n-1}|\beta_n|^2$ with $|\beta_n|^2\geq0$.

We call these particular states as ``reserved'' states in this paper. Given $|\alpha_{n+1}|^2=1$ or $\Omega_{n+1}\tau=\pi$, it is immediately to find that the measurement interval for the first reserved state $n=n_r^{(1)}$ can be written as
\begin{equation}\label{tau}
\tau=\tau_r=\frac{\pi}{\Omega_{n_r^{(1)}+1}}=\frac{\pi}{\sqrt{g^2\left[n_r^{(1)}+1\right]+\Delta^2/4}}
\end{equation}
or its multiple. Conversely, for a given measurement interval $\tau_r$, we have a group of reserved states:
\begin{equation}\label{RsvStates}
n_r^{(k)}=n_r^{(1)}+(k^2-1)\left[n_r^{(1)}+\delta^2+1\right],
\end{equation}
where $k\in\mathbb{N}_+$ and $\delta^2=\Delta^2/(4g^2)$. The reserved states are then not unique for $\Omega_{n_r^{(k)}+1}\tau_r=k\pi$ with $k$ integer. Also they are under protection and the populations would be gradually concentrated around them by unconditional measurements. Of course $n_r^{(k)}$ should be understood as the closest integers to the right hand of Eq.~(\ref{RsvStates}) and $k$ indicates the order of the reserved states.

\begin{figure}[htbp]
\centering
\includegraphics[width=0.95\linewidth]{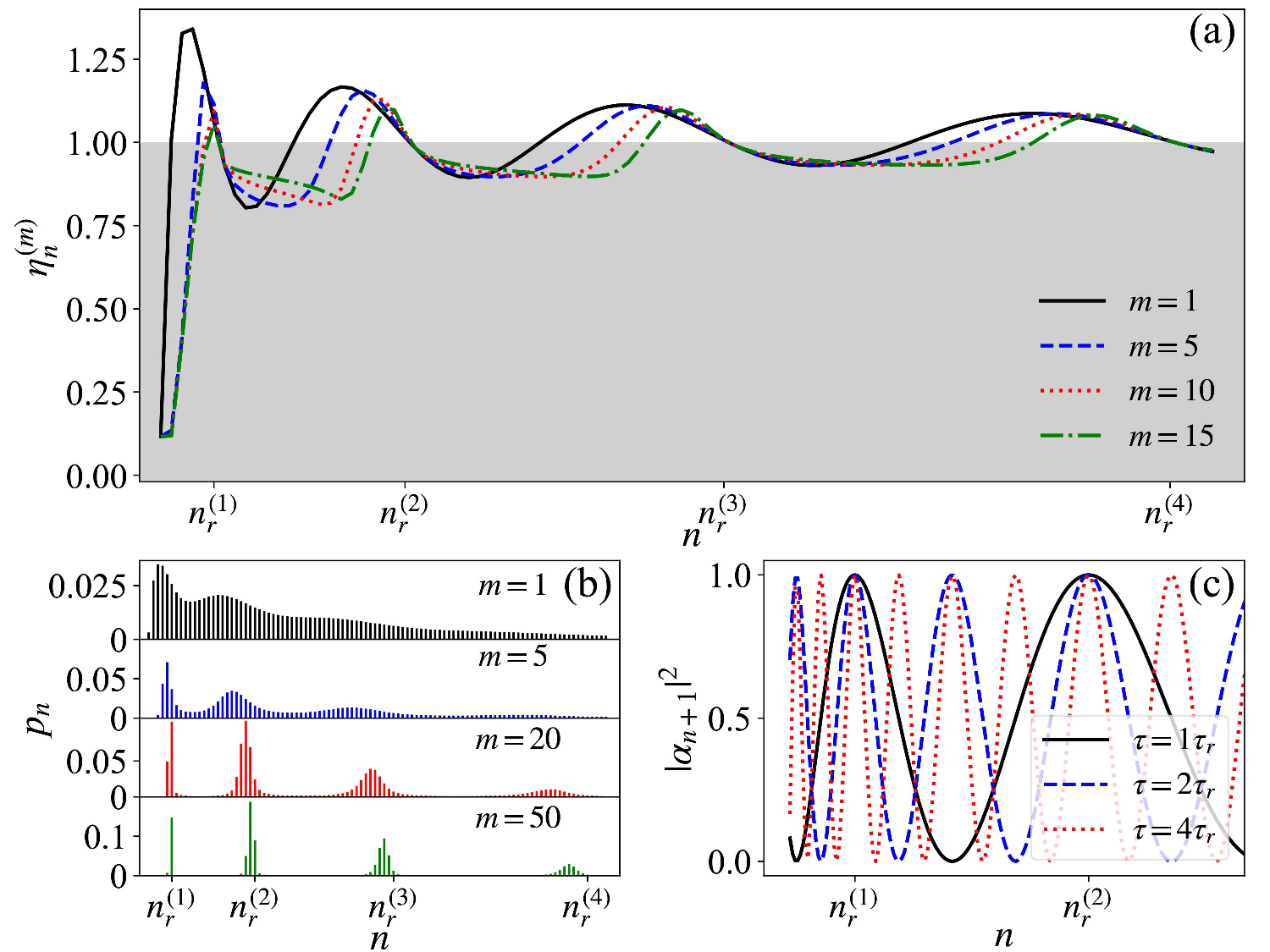}
\caption{(a) Changing ratio of population $\eta$ as a function of the Fock-state index $n$ after various numbers of measurements under a fixed measurement interval $\tau=\tau_r$ in Eq.~(\ref{tau}) with a given reserved state $n_r^{(1)}=5$. Grey area is upper bounded by $\eta=1$, representing the regions where the populations on $|n\rangle$ keep falling during the unconditional measurements. (b) Population histograms in the Fock space of the resonator under various numbers of measurements. For (a) and (b) the resonator is initialized as a thermal state with a high temperature $T=1$ K to have a wide range of Fock states with non-negligible populations. (c) Cooling coefficient $|\alpha_{n+1}|^2$ as a function of $n$ under various $\tau$. The coupling strength and the detuning are set as $g=0.04\omega_b$ and $\Delta=0.02\omega_b$, respectively. With $n_r^{(1)}=5$, we have $n_r^{(2)}\approx23$, $n_r^{(3)}\approx53$, and $n_r^{(4)}\approx95$ due to Eq.~(\ref{RsvStates}). }\label{Coefficient}
\end{figure}

In Fig.~\ref{Coefficient}(a), the population changing ratio
\begin{equation}
\eta^{(m)}_n=\frac{p_n^{(m)}}{p_n^{(m-1)}}
\end{equation}
is plotted to exhibit the effect of unconditional measurements with a fixed interval $\tau_r$ in Eq.~(\ref{tau}), where $p_n^{(m)}$ is the population on $|n\rangle$ after $m$ measurements. The population $p_n$ grows with measurements when $\eta^{(m)}_n>1$ (the white area); and declines when $\eta^{(m)}_n<1$ (the grey area). For any $m$, the population-changing ratios manifest similar pattern in the Fock bases. A minimal value of $\eta$ remains on the ground state $|n=0\rangle$, which is found to be a constant $\eta_0=|\alpha_0|^2\approx0.12$ due to Eq.~(\ref{pn}). For $n>0$, $\eta_n^{(m)}$ increases rapidly with $n$, approaches a local peak value, and then drops to unit nearby the first reserved state $n_r^{(1)}$. The Fock index of the peak value moves towards $n_r^{(1)}$ under repeated unconditional measurements on qubit, which generate a significant concentration of the populations over a low-energy range $[0\sim n_r^{(1)}]$ onto the first reserved state $|n_r^{(1)}\rangle$. The range of $\eta^{(m)}_n>1$ contracts with $m$, yet always covers the proximity of $|n_r^{(1)}\rangle$. When $n$ becomes larger than $n_r^{(1)}$, the population-changing ratio drops below unit until the next reserved state $n_r^{(2)}$. For a higher temperature resonator, that has non-negligible populations over a wider range of Fock states, one can see more separable ranges of states with $\eta^{(m)}_n>1$. As measurements are repeatedly implemented, all of these bulges are contracting and moving towards the reserved states $n_r^{(k)}$ given by Eq.~(\ref{RsvStates}), $k\geq1$. The thermal-distributed populations would then be gradually concentrated to the reserved states. And more measurements or a longer running time are required to enhance the populations on the higher-order reserved states.

We use the population histograms for the resonator on Fock bases in Fig.~\ref{Coefficient}(b) to demonstrate the population concentration under repeated unconditional measurements. It is interesting to see that the exponential-decay distribution for the thermal state is dramatically reshaped by measurements. The reserved states are then distinguished by collecting more and more populations and clearly $p_{n_r^{(1)}}$ dominates in the low energy scale. It is therefore instructive to search an efficient way to generate a high-fidelity Fock state $|n_r^{(1)}\rangle$ from a mixed state, especially from a thermal state which is maximally populated on the ground state. That constitutes the main target of the first step in our cooling protocol. And the rest question in this step is how to suppress the populations over the high-order reserved states $n_r^{(k)}$, $k\geq2$, which are also under protection and even get more occupied. To avoid their disturbance, one can choose a proper reserved state $|n_r^{(1)}\rangle$ for the resonator under a given initial temperature. For the thermal state with an average occupation $\bar{n}_{\rm th}$, its root-mean-square deviation is $\Delta n=(\bar{n}_{\rm th}+\bar{n}_{\rm th}^2)^{1/2}$, and the accumulated population up to $|n=N\rangle$ reads
\begin{equation}
P(N)=\frac{1}{\bar{n}_{\rm th}}\sum_{n=0}^N\left(\frac{\bar{n}_{\rm th}}{1+\bar{n}_{\rm th}}\right)^n=1-\left(\frac{\bar{n}_{\rm th}}{1+\bar{n}_{\rm th}}\right)^N.
\end{equation}
It is immediately found that $P(N=\bar{n}_{\rm th}+4\Delta n)\geq0.99$. Then to prevent the population concentration on the second reserved state, it is required that
\begin{equation}\label{Condition}
n_r^{(2)}\geq\bar{n}_{\rm th}+4\Delta n,
\end{equation}
by which the original thermal populations around $|n_r^{(2)}\rangle$ become ignorable. According to Eq.~(\ref{RsvStates}), this condition essentially sets a lower bound for the first reserved state $n_r^{(1)}$:
\begin{equation}\label{ConditionRsvState}
n_r^{(1)}\geq\frac{1}{4}\left[\bar{n}_{\rm th}-3(\delta^2+1)\right]+\Delta n.
\end{equation}
It can be applied to choose the reserved state for a high fidelity Fock-state preparation.

According to Eq.~(\ref{AlphaBeta}), both the cooling coefficient $|\alpha_{n+1}|^2$ and the reserved state $n_r^{(1)}$ are significantly influenced by the measurement interval $\tau$. Note $\tau$ could be a multiple of $\tau_r$ in Eq.~(\ref{tau}). Given a target state $|n_r^{(1)}\rangle$, Fig.~\ref{Coefficient}(c) demonstrates $|\alpha_{n+1}|^2$ as a function of the Fock-state index with various measurement intervals. Using the shortest interval $\tau/\tau_r=1$, we can reduce the unwanted populations of a wide range bounded by $|n_r^{(1)}\rangle$ and $|n_r^{(2)}\rangle$. Using a longer one, the period of $|\alpha_{n+1}|^2$ is reduced, indicating a sharper population distribution around the reserved states due to Eq.~(\ref{pn}). A Fock state with a higher fidelity is crucial for realizing the ground-state cooling with a larger probability. Thus an unequal-spacing sequence of $M$ unconditional measurements can be constructed by setting $\tau_i=j\tau_r$ with an optimized integer $j$ for the $i$th measurement. $i$ runs from $1$ to $M$. If $j\in\{1,2,\cdots,d\}$, we then have to test $d^M$ sequences by brute force.

In Sec.~\ref{UnitProbCoolingSec}, we apply a reinforcement learning method to quickly generate an optimized sequence of time spacings for unconditional measurements (as an input of parameters for the implementation of our protocol), yielding an almost complete population concentration on the reserved state. An intelligent agent in the reinforcement learning would take actions according to the current status (population distribution) and then update its experience depending on rewards or punishments through a feedback mechanism. It is capable to generate a sequence of optimal actions to achieve a certain target (reserved-Fock-state fidelity) by trial and error in computer simulation.

\subsection{Population transfer based on conditional measurement}

In contrast to the unconditional measurement, its conditional counterpart discards the system population over unprojected subspaces, that yields nondeterministic rounds of cooling. Here we employ a projective operator $M_e=|e\rangle\langle e|$ based on the qubit excited state. Rather than the projection $M_g=|g\rangle\langle g|$ for conventional cooling-by-measurement, $M_e$ gives rise to a nondeterministic POVM for transferring population of the resonator from the higher-level states to the lower ones.

In particular, the qubit is prepared at the ground state. After performing the conditional measurement in the end of the joint evolution of the composite system lasting $\tau$, the resonator state becomes
\begin{equation}
\rho_b(t+\tau)=\frac{\langle e|U(\tau)\rho_b(t)\otimes|g\rangle\langle g|U^\dagger(\tau)|e\rangle}{P_s},
\end{equation}
where
\begin{equation}
P_s={\rm Tr}[R(\tau)\rho_b(t)R^\dagger(\tau)]
\end{equation}
represents the measurement probability and
\begin{equation}\label{Kraus}
R(\tau)\equiv\langle e|U(\tau)|g\rangle=\sum_{n=1}\frac{-ie^{i\Delta\tau}g\sqrt{n}\sin\Omega_n\tau}{\Omega_n}|n-1\rangle\langle n|
\end{equation}
is the Kraus operator defined in the Hilbert space of the resonator. According to the Naimark's dilation theorem~\cite{Naimark}, the projective measurements performed on the ancillary qubit induce POVMs $\mathcal{M}(\tau)[\mathcal{O}]\equiv R(\tau)\mathcal{O} R^{\dagger}(\tau)$ acted on the resonator. Then the resonator state (without normalization) after a single conditional measurement reads
\begin{equation}\label{RhobAfterCM}
\rho_b(t+\tau)=\mathcal{M}(\tau)[\rho_b(t)]=\sum_{n=1}|\beta_n|^2p_n|n-1\rangle\langle n-1|.
\end{equation}
Equation~(\ref{RhobAfterCM}) describes the downward population transfer between neighboring pairs of states $p_{n-1}\leftarrow|\beta_n|^2p_n$ with an $n$-dependent factor $|\beta_n|^2$ given by Eq.~(\ref{AlphaBeta}). The transfer efficiency could be fully attained up to $|\beta_n|^2=1$ by choosing a proper measurement time spacing $\tau$. In this situation, the Kraus operator acts the same as a lowering operator $R_n\sim|n-1\rangle\langle n|$, completely transferring the population from $|n\rangle$ to $|n-1\rangle$ for a given $n$, if the resonator has been prepared as $|n\rangle$ in advance. Then optimized measurement intervals allow projective measurements to bring the whole population from a high-level Fock state to the ground state step by step. After the first step of unconditional measurements, we have $n=n_r$ (the superscript of the first reserved state is omitted for simplicity in what follows) and then the optimal measurement interval of the first round of the conditional measurements is found to be
\begin{equation}\label{OptTau}
\tau_{\rm opt}=\frac{\pi}{2\Omega_{n_r}}.
\end{equation}
by the condition of $|\sin(\Omega_{n_r}\tau)|=1$ due to Eq.~(\ref{AlphaBeta}). And in the following $k$th round, $n_r$ is updated to $n_r-(k-1)$, $k\geq2$.

One can find from Eq.~(\ref{OptTau}) that more measurements are demanded for a reserved state with a larger $n_r$ and $\tau_{\rm opt}$ becomes longer as a result of slower transitions in the subspaces with a smaller number of excitations. As conditional measurements are implemented, the resonator evolution is conditional on the strongly correlated outcomes and the consequence of measurements forms a conditional trajectory~\cite{SpinBathPurification} in the parameter space. With the optimized period $\tau_{\rm opt}$, a wanted measurement result that the qubit is in its excited state suggests that one unit of energy has been faithfully extracted from the resonator. Therefore the success probability does not significantly decay under those particular POVMs and the energy is constantly leaking outside until the resonator is found at the ground state.

\section{Efficient Cooling with near-unit probability}\label{UnitProbCoolingSec}

\begin{figure*}[htbp]
\centering
\includegraphics[width=0.7\linewidth]{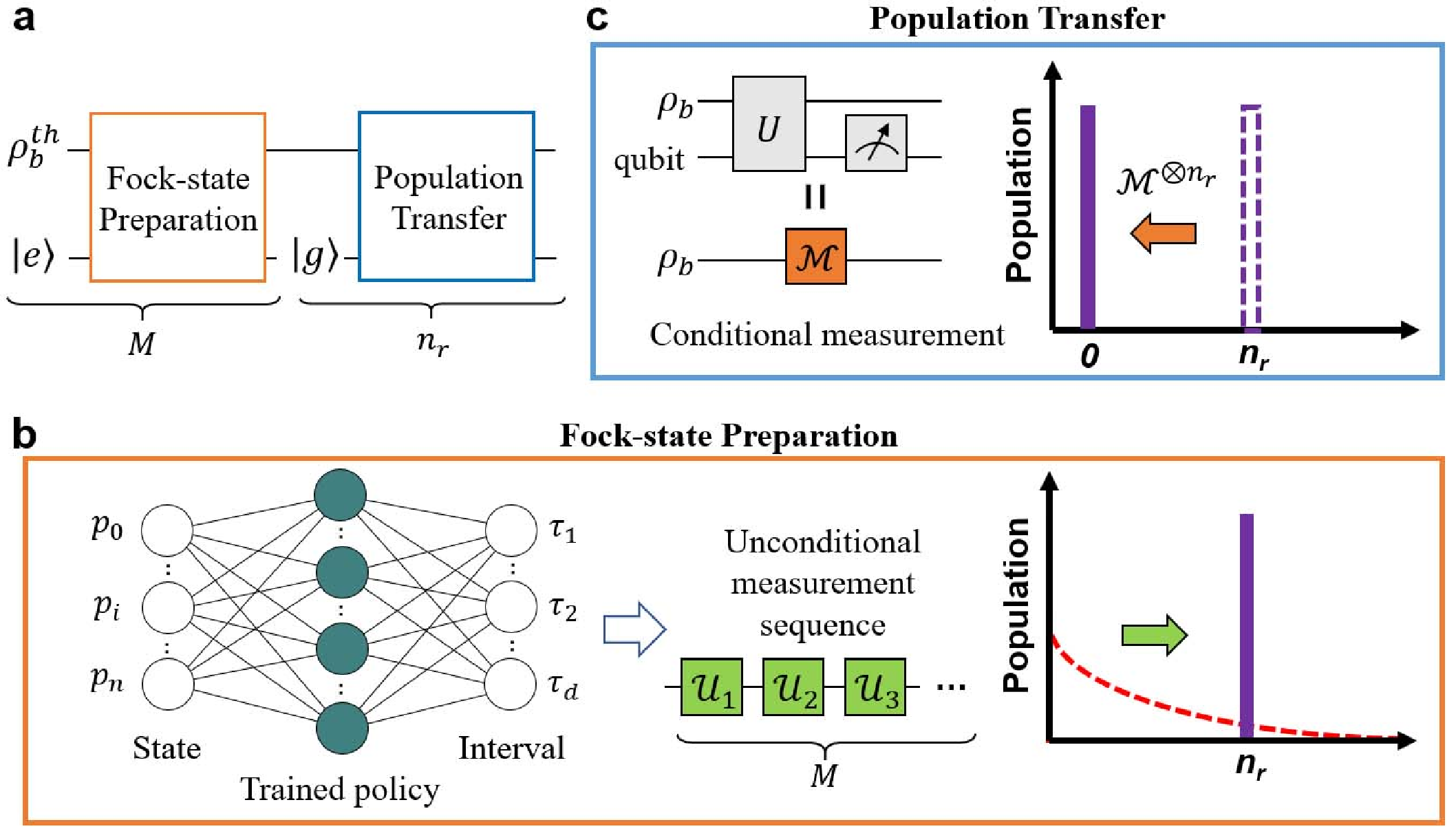}
\caption{(a) Framework of our near-unit-probability cooling protocol consisting of two steps. The model consists of a to-be-cooled resonator $\rho_b$ initially at a thermal state and an ancillary qubit prepared at the excited state $|e\rangle$ and the ground state $|g\rangle$ during the first and second steps, respectively. (b) Diagram of the Fock-state-preparation step assisted by reinforcement learning. A policy constructed by neural networks is trained to choose optimal measurement interval based on the current state, where the input of the neural networks is the resonator populations $s=\{p_0, p_1, \cdots, p_n\}$ and the output is a probability distribution of various unconditional measurement intervals. After training, an optimized sequence of measurement intervals is generated for unconditional measurements. During this step, the resonator can be efficiently reshaped from a thermal state to a reserved Fock state $|n_r\rangle$. (c) Diagram of the population-transfer step. After a period of joint unitary evolution, a projective measurement implemented on the ancillary qubit gives rise to a POVM on the resonator, which could transfer population from a higher Fock state to a lower one. After $n_r$ rounds of conditional measurements, the population over the reserved state is fully transferred to the ground state.}\label{Model}
\end{figure*}

Combining the preceding unconditional and conditional measurements, we are on the stage to present our two-step cooling protocol as shown in Fig.~\ref{Model}(a). In the first step (see the orange frame), the ancillary qubit is prepared as the excited state and the initially thermal resonator is reshaped to be the first reserved Fock state $|n_r\rangle$ by unconditional measurements $\mathcal{U}(\tau_i)$ in Eq.~(\ref{mathcalU}), where $\tau_i$ represents the time interval of the $i$th round of evolution and measurement. As we have analyzed in Sec.~\ref{Fockstate}, a shorter $\tau_i$ is helpful to suppress the populations on the unwanted states but is inefficient to achieve a high-fidelity Fock state $|n_r\rangle$. In contrast, a longer $\tau_i$ accelerates the Fock-state generation but might leave more populations on the high-order reserved states. We use a reinforcement learning method to find an optimized and finite sequence of unconditional measurements with varying intervals to achieve a Fock state with a high fidelity.

We offer an action set or space $\tau\in\{\tau_1, \tau_2, \cdots, \tau_d\}$ in Fig.~\ref{Model}(b) by a policy neural network, where $\tau_j=j\tau_r$ indicate various measurement intervals and $d$ is the set size. Aiming at a Fock-state fidelity as high as possible, the policy neural network is trained to learn a sequence of unequal measurement intervals when implementing the unconditional measurements on ancillary qubit. A distributed proximal policy optimization algorithm~\cite{DPPO} is employed for optimization and more details could be found in Appendix~\ref{DPPOSec}. Note that the sequence of measurement intervals generated by reinforcement learning serves as a parametric input for implementing the unconditional measurements. Once the reserved Fock state $|n_r\rangle$ is prepared by $M$ rounds of unconditional measurements, it is loaded to the second step (see the blue frame). $M$ is determined during a pre-training process, which is an adjustable parameter to ensure a sufficiently high population concentration on the reserved state and to avoid vain measurements. As demonstrated in Fig.~\ref{Model}(c), the ancillary qubit is flipped to the ground state in the second step and the projective measurements $M_e=|e\rangle\langle e|$ are performed on the qubit with joint evolutions of varying intervals $\tau_{\rm opt}$ in Eq.~(\ref{OptTau}). It induces a POVM $\mathcal{M}$ on the resonator capable of transmitting the population on $|n\rangle$ to $|n-1\rangle$ with a near-unit probability. Then after extra $n_r$ rounds of conditional measurements, the resonator is cooled down to the ground state.

\begin{figure}[htbp]
\centering
\includegraphics[width=0.95\linewidth]{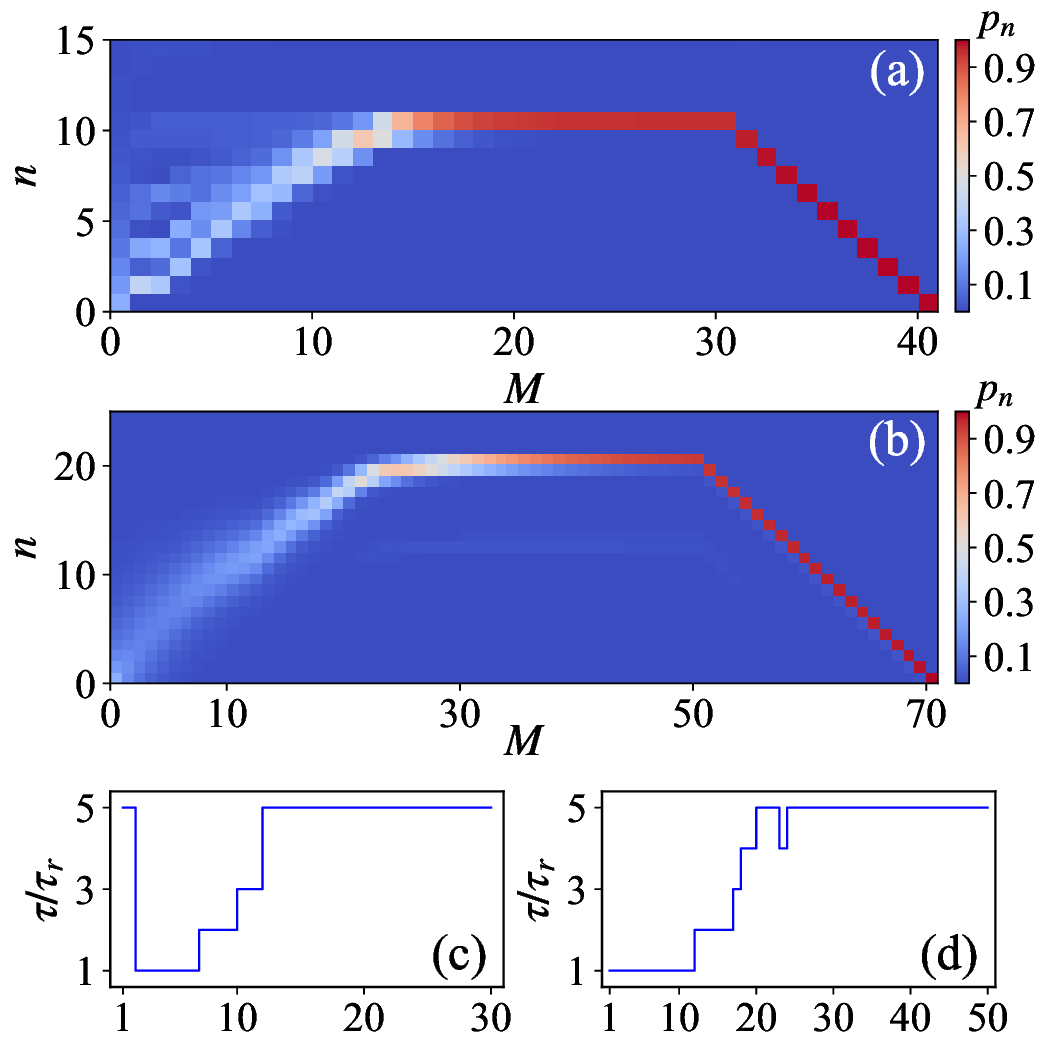}
\caption{(a) and (b): Population distributions of the resonator in the Fock space as a function of the measurement number. Starting from a thermal state, the populations across a wide range are gradually concentrated to the reserved state (a) $|n_r=10\rangle$ by $30$ rounds of unconditional measurements and (b) $|n_r=20\rangle$ by $50$ rounds of unconditional measurements, respectively. After that, $n_r$ rounds of conditional measurements are performed on the qubit to transfer the accumulated populations on $|n_r\rangle$ to the ground state in a stepwise way. (c) and (d): Optimized unconditional measurement sequences for the reserved states (a) $n_r=10$ and (b) $n_r=20$, respectively, under the same action space $\tau\in\{\tau_1, \tau_2, \cdots, \tau_5\}$. The resonator frequency is set as $\omega_b=3.7$ GHz and the initial temperature is $T=0.1$ K. $\Delta=0.05g$ and $g=0.04\omega_b$.}\label{States}
\end{figure}

Our two-step protocol could be applied to cool down a nanomechanical oscillator in gigahertz~\cite{NVcentreResonator,Nonomechanical,MechanicalResonator}, whose eigenfrequency is $\omega_b=3.7$ GHz. The coupling strength between the resonator and the ancillary qubit is $g/\omega_b=0.04$ and the initial temperature is $T=0.1$ K. In Figs.~\ref{States}(a) and \ref{States}(b), the time-evolved population distributions of the resonator, i.e., the vertical ordered histograms, are plotted under various number of measurements (including both unconditional and conditional measurements). The data for $M=0$ describe the initial thermal state. In Fig.~\ref{States}(a), as implemented by unconditional measurements, the population-transfer ratio $\eta_{n}$ on the reserved state $|n_r=10\rangle$ is found to be always greater than unit. The populations over the lower-energy levels (especially the ground state) are gradually collected to the higher levels until the reserved state. After $M=15$ unconditional measurements, $|n_r\rangle$ has been distinguished with a fidelity $F_r\equiv\langle n_r|\rho_b|n_r\rangle$ over $0.68$. For $M=20$ and $M=30$, its fidelity is over $0.94$ and close to unit, respectively. The Fock-state preparation cannot be trivially regarded as a consequence of energy swap between the ancillary qubit and the target resonator, where no postselection occurs on the qubit state. In addition, the population on the reserved state cannot be monotonously enhanced without the time-spacing optimization, although the energy gain of the resonator attributes to the excitation of the ancillary qubit. When the Fock state $|n_r\rangle$ is prepared, the second step with the conditional measurements starts to transfer its population back to the ground state $|0\rangle$. In the end of the whole cooling process, the fidelity of the ground state reaches $0.999997$ with a success probability over $P_s=95\%$.

In Fig.~\ref{States}(b), more unconditional measurements are required to achieve the larger target reserved state $|n_r=20\rangle$ due to the higher energy level and a lower initial population on the reserved state. After $M=30$ measurements, the neighboring state $|n=19\rangle$ is significantly populated with $\langle n=19|\rho_b|n=19\rangle=0.35$. After $M=50$ measurements, the fidelity of the reserved state becomes over $F_r=0.94$. Similarly, $n_r=20$ rounds of conditional measurements are implemented on the qubits following the Fock-state preparation procedure and the final fidelity of the ground state reaches $0.999996$ with a success probability over $P_s=94\%$. In the cases of $n_r=10$ and $n_r=20$, the average occupation number $\bar{n}={\rm Tr}[\hat{n}\rho_b]$ is reduced from the initial value $\bar{n}\approx3.06$ to $\bar{n}=1.57\times10^{-5}$ and $\bar{n}=3.83\times10^{-6}$, respectively, by over five orders in magnitude.

In the existing nondeterministic cooling protocols~\cite{Li2011,Purification}, a product ground state of resonator and qubit $|g0\rangle$ is decoupled from the other subspaces by rounds of postselections based on direct projective measurements. The non-negligible population distributed over the unwanted excited states yields a low success probability for cooling. In our protocol, however, the target system has been already purified as a Fock state $|n_r\rangle$ and is then subject to the nondeterministic POVM described by Eq.~(\ref{Kraus}) based on the projective measurements. Thus there is little loss in population during the second step.

The resource cost of our protocol is also limited. During the first step for Fock-state preparation by unconditional measurements, there is no postselection over the ancillary qubit. The success probability is thus unit in principle. During the second step for population transfer, the ancillary qubit is prepared as the ground state and projected to the excited state in each round of the conditional measurements [see Fig.~\ref{Model}(a)]. So that it is possible to require more than one qubit to compensate the finite probability of postselection. For the result in Fig.~\ref{States}(a), the success probability is about $P_s\approx95\%$, which means merely less than $5$ out of every $100$ experiments (suppose that we have an ensemble of the same system and perform the measurement sequence many times) will fail to cool down the resonator.

Figures~\ref{States}(c) and \ref{States}(d) demonstrate the time-spacing sequences in the unconditional-measurement step, which are generated by the well-trained policy with reserved states $|n_r=10\rangle$ and $|n_r=20\rangle$, respectively. The key strategies learned by the policy seem to share some similarities. At the first several rounds, they prefer to use more shorter intervals to collect more populations around the first reserved state. Effectively it reduces the populations held by the higher-order reserved states. Then they choose gradually longer measurement-intervals and in the last few rounds stick to the maximum value. This strategy is beneficial to sharpen the population distribution around the reserved state, acting as a fine manipulation for purification.

\section{Cooling under thermal bath}\label{discussion}

It is inevitable to study the cooling process in an open-quantum-system scenario by considering decoherence of the target resonator, which arises from the interaction between resonator and the external bath. The cooling efficiency is expected to decline in the presence of a thermal bath with a finite temperature. In this case, the system evolution intersected by the measurements can be simulated by the master equation
\begin{equation}\label{ME}
\dot{\rho}(t)=-i[H, \rho(t)]+\gamma(\bar{n}_{\rm th}+1)\mathcal{D}[a]\rho(t)+\gamma\bar{n}_{\rm th}\mathcal{D}[a^\dagger]\rho(t),
\end{equation}
where $\rho(t)$ represents the composite system of the resonator and the qubit, $\gamma$ is the decoherence rate, and $\mathcal{D}[A]$ represents the Lindblad superoperator
\begin{equation}\label{Lindblad}
\mathcal{D}[A]\rho(t)\equiv A\rho(t)A^\dagger-\frac{1}{2}\left\{A^\dagger A, \rho(t)\right\}.
\end{equation}
To distinguish the decoherence effect from the thermal bath, we perform the same unconditional measurement sequences as trained by reinforcement learning for the closed-quantum-system scenario in Sec.~\ref{UnitProbCoolingSec} and use the same optimal measurement intervals in the conditional-measurement step as given by Eq.~(\ref{OptTau}).

\begin{figure}[htbp]
\centering
\includegraphics[width=1\linewidth]{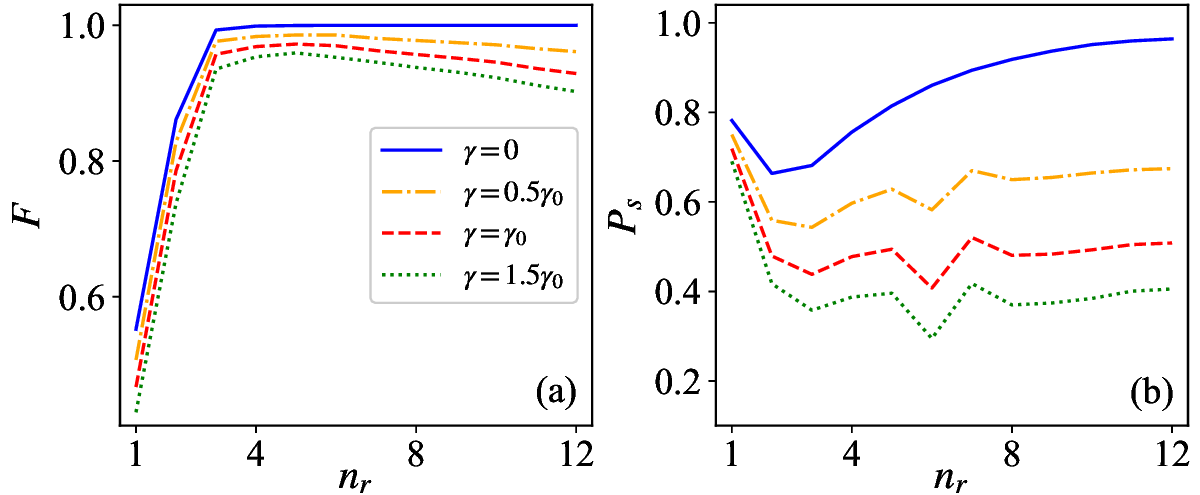}
\caption{(a) Ground-state fidelity $F$ in the end of the cooling protocol and (b) Success probability $P_s$ of the resonator as functions of the first reserved state under various decoherence rates. The unit of decoherence rate is chosen as an experimental-relevant value $\gamma_0=10^{-5}\omega_b$~\cite{Nonomechanical}. The other parameters are the same as Fig.~\ref{States}.}\label{FidelityAndProbs}
\end{figure}

In Fig.~\ref{FidelityAndProbs}(a), the fidelity of the target resonator's state $F\equiv\langle0|\rho_b|0\rangle$ in the end of the cooling with respect to its ground state is shown as a function of the reserved state $|n_r\rangle$ under various decoherence rates. For the reserved states lower than the lower-bound, which is found to be $\sim 3$ by Eq.~(\ref{ConditionRsvState}) using the parameters in plotting, the non-negligible population accumulated on the high-order reserved states would remarkably reduce the purity of the target Fock state. For $n_r<3$, the fidelity is lower than $0.87$ even in the decoherence-free case, signifying an inefficient cooling. In sharp contrast, one can see for $n_r\geq3$ that the resonator can be cooled down to the ground state with a near-unit fidelity when $\gamma=0$. Although the fidelity declines slowly with increasing $\gamma$, it is maintained about $0.90$ when $\gamma=1.5\gamma_0$ for $n_r=12$. The ground-state fidelity thus manifests robustness against the thermal bath. For $n_r\geq3$, the fidelity of a larger Fock state is more suppressed in the presence of a thermal bath than a smaller one. It arises from the fact that a higher reserved state indicates more rounds of conditional measurements as well as a much longer running time in the second step of our protocol for population transfer. And by the master equation~(\ref{Lindblad}), the effective decay rate is proportional to the initial average population of the resonator. Then a higher-level state is more sensitive to a thermal bath than a lower-level one.

The pattern of the success probability $P_s$ is shown in Fig.~\ref{FidelityAndProbs}(b), which is also roughly seperated by the lower bound $n_r\sim3$. For $n_r\leq3$, the success probability of $n_r=1$ is higher than those of $n_r=2$ and $n_r=3$. When $\gamma=0$, it is close to $80\%$. That is understandable since there is only one projection in the population-transfer step yet it does not make much sense in cooling since the corresponding ground-state fidelity is less than $0.5$ as shown in Fig~\ref{FidelityAndProbs}(a). For $n_r\geq3$, the success probability is roughly enhanced with increasing $n_r$ and gradually decays with increasing $\gamma$. In the absence of the thermal bath, the success probability reaches $P_s=96\%$ for $n_r=12$ and it can be maintained over $40\%$ even for $\gamma=1.5\gamma_0$, which is still much higher than that of the cooling protocols relying entirely on the optimized conditional measurements~\cite{TwoModeCooling}.

\section{Discussion and Conclusion}\label{ConclusionSec}

In summary, we have proposed a two-step cooling protocol featured with both high efficiency and a near-unit success probability. It is applied to cooling a thermal resonator down to its ground state by coupling to an ancillary qubit under measurement. The first step consisting of unconditional measurements is in charge of transforming the target resonator from a mixed state, e.g., a thermal state, to a reserved Fock state. The time-spacing sequence of the evolution-and-measurement rounds can be optimized in advance by the distributed proximal policy optimization algorithm in reinforcement learning, to compromise the effects from various measurement intervals on purifying the reserved state and suppressing the populations on the unwanted states. For a proper reserved state, the target resonator would be prepared as a Fock state with a near-unit fidelity by dozens of measurements. The second step relies on a nondeterministic POVM induced by the projection on the excited state of the ancillary qubit that is prepared at the ground state after each round of joint evolution of resonator and qubit. The POVMs spaced by an updated optimized measurement interval would move downwards the populations on the reserved state with an almost unit probability. By stepwisely extracting the energy from resonator, the prepared reserved state approaches the ground state.

In contrast to the existing cooling protocols based entirely on the projective measurements that is nondeterministic for every round, the current protocol hybridizes the determinacy of unconditional measurements and the high efficiency on population reduction of conditional measurements. Then the cooling rate is maintained at a high level yet without much loss of the success probability caused by postselections. Our protocol does not require the initial nonvanishing occupation on the ground state for the resonator. It paves a practical revenue of cooling by measurement since the initial thermal average occupation of the resonator is reduced by five orders in magnitude with a success probability close to unit. And in the same time, the experimental cost is moderate since one uses only dozens of measurements. Also it offers a promising example for an interdisciplinary application of quantum state manipulation and machine learning for optimization.

\section*{Acknowledgements}
We acknowledge financial support from the National Natural Science Foundation of China (Grant No. 11974311).

\appendix
\section{Distributed Proximal Policy Optimization}\label{DPPOSec}

\begin{figure}[htbp]
\centering
\includegraphics[width=0.95\linewidth]{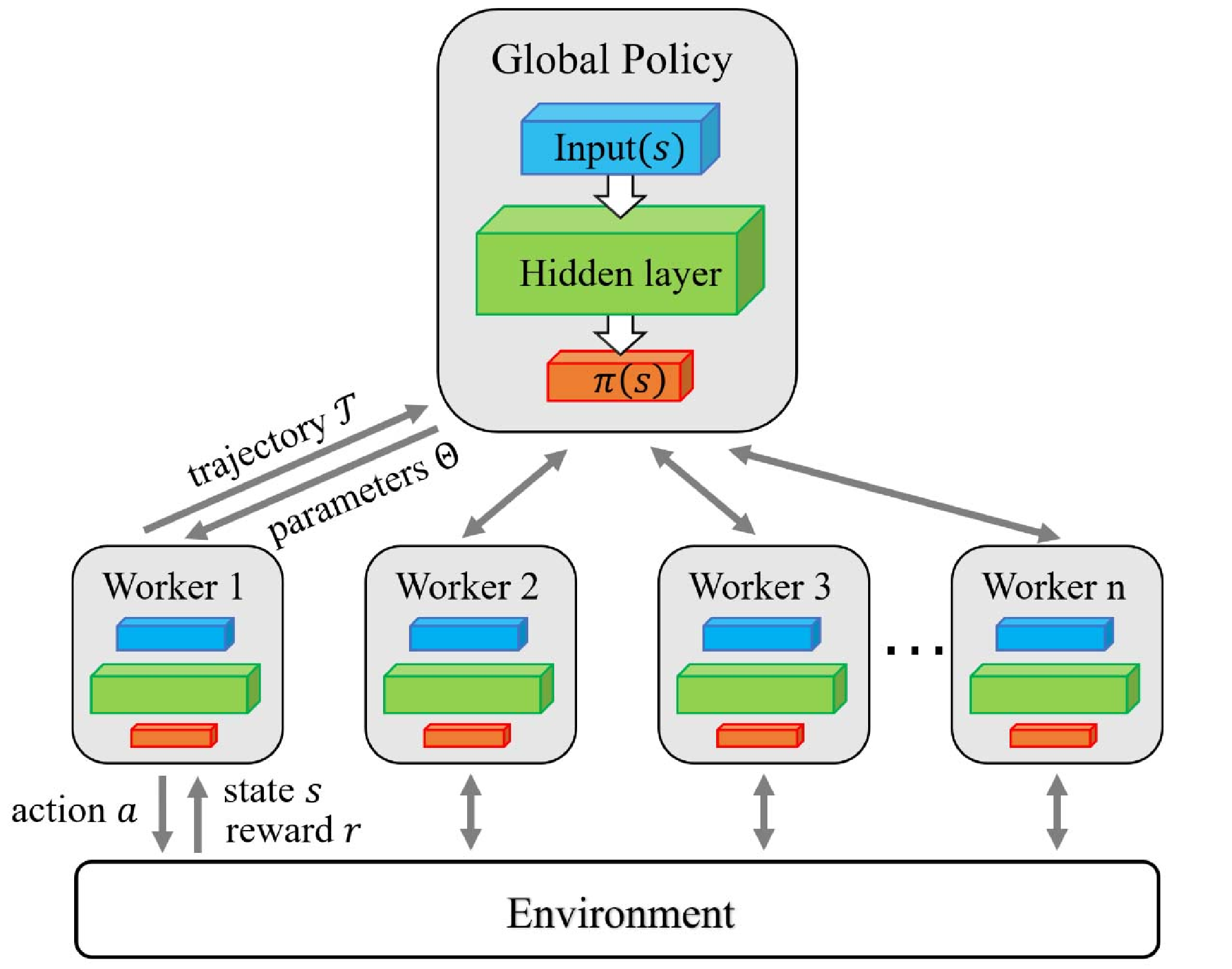}
\caption{Diagram of the policy updating in distributed proximal policy optimization algorithm.}\label{DPPOFig}
\end{figure}

This appendix is devoted to reveal more details about the distributed proximal policy optimization (DPPO) used to generate an optimized unconditional measurement sequence in Sec.~\ref{UnitProbCoolingSec}. DPPO algorithm is a distributed variant of proximal policy optimization (PPO)~\cite{DPPO}, in which an updatable policy is trained as an actor to choose the comparatively optimized or correct actions toward the final goal and a critic is trained to evaluate quantitatively if the actions chosen by the policy should be encouraged. In the conventional PPO, there are two policies and one critic. Both policy and critic are constructed by neural networks with individual sets of parameters. The two policies share the same neural network structure. The old policy is responsible for collecting data by interacting with an environment and the new one would use data collected by the old policy to update its network. In DPPO, there is a global policy and several worker policies. All policies have the same network construction. Computation is distributed over parallel instances of policy and environment, and data collection is done by several parallel threads as shown in Fig.~\ref{DPPOFig}. In each thread, there is a worker policy interacting independently with the environment. At the first trial, an individual worker policy chooses an action $a_1$ according to the initial state $s_0$, then in the environment the action is taken and consequently the state is modified to be $s_1$. The environment would return a reward $r_1$ based on a well-defined reward function, after which both the updated state $s_1$ and the reward $r_1$ are returned to the worker policy. The interaction is repeated for several times until a trajectory is completed $\mathcal{T}=\{s_0, a_1, r_1, s_1, \cdots\}$. For updating the global policy, a batch of trajectories are required to be collected, so distributing the collection task over parallel threads can remarkably speed up the training process. Note that in between the data collection and the global policy updating, the network parameters $\Theta$ of all worker policies should be timely updated, ensuring the global policy is always one version ahead of all worker policies.

As to our two-step cooling protocol, the input states of policies are populations of the density matrix of the target resonator $s_i=\{p_0, p_1, p_2, \cdots, p_{n_c}\}$ with a cutoff $n_c$ in the Fock space. The dimension of the action space is set as five: $a\in\{1, 2, 3, 4, 5\}$, representing the multiplier $j$ of the measurement interval $\tau_r$ given by Eq.~(\ref{tau}) for the first reserved state. The ``environment'' (not the thermal bath for open quantum systems) would implement unconditional measurements with varying intervals according to actions chosen by policies. The reward is a tangent function $r(s_i, a_i)=10\tan{(F_r\pi/2)}$ of the reserved state fidelity $F_r$, which encourages the fidelity to approach unit as close as possible. After training, the global policy is able to generate an optimal sequence $\mathcal{S}_{\rm opt}=\{\tau_1, \tau_2, \cdots, \tau_M\}$ consisting of optimized intervals for unconditional measurements, where $M$ is convergent for a high population concentration on the reserved state.

The reinforcement learning method is much more efficient in searching the optimized interval sequences than the brute-force searching, which will cost an exponential-increasing resource in both calculation time and memory space. In the context of Fig.~\ref{States}(a), there are $d=5$ options of measurement-interval for each round and the unconditional measurement sequence consists of $M=30$ rounds. So that there are $5^{30}$ kinds of arrangements. It takes about $0.15$ seconds to run one sequence of measurements by a regular personal computer (Intel Core i7-9700 processor $3.00$ GHz and memory $6$ GB in our numerical simulation). In contrast, our reinforcement learning accelerated by distributed sampling over $4$ threads in DPPO takes about $1$ hour to find an optimized arrangement, demonstrating an ultra advantage.

\bibliographystyle{apsrevlong}
\bibliography{ref}

\end{document}